\documentstyle[prb,aps]{revtex}

\begin{document}
\draft
\twocolumn[\hsize\textwidth\columnwidth\hsize\csname
@twocolumnfalse\endcsname

\title{Dynamics and Scaling of One Dimensional Surface Structures}
\author{Navot Israeli$^1$, Hyeong-Chai Jeong$^2$, Daniel Kandel$^1$ and
John D. Weeks$^3$}
\address{$^{1}$Department of Physics of Complex Systems,\\ Weizmann
Institute of
Science,
Rehovot 76100, Israel}
\address{$^2$Department of Physics, Sejong University,\\
Kwangjinku, Seoul 143-747, Korea}
\address{$^{3}$Institute for Physical Science and Technology and\\
Department of
Chemistry,\\
University of Maryland, College Park, Maryland 20742\\
}
\maketitle

\begin{abstract}
We study several one dimensional step flow models. Numerical simulations
show that the slope of the profile exhibits scaling in all cases. We
apply a
scaling ansatz to the various step flow models and investigate their
long
time evolution. This evolution is described in terms of a {\em
continuous}
step density function, which scales in time according to 
$D(x,t)=F(xt^{-1/\gamma })$. The value of the scaling exponent $\gamma$
depends on the mass transport mechanism. When steps exchange atoms with
a
global reservoir the value of $\gamma $ is 2. On the other hand, when
the
steps can only exchange atoms with neighboring terraces, $\gamma=4$. We
compute the step density scaling function for three different profiles
for
both global and local exchange mechanisms. The computed density
functions
coincide with simulations of the discrete systems. These results are
compared to those given by the continuum approach of Mullins.
\end{abstract}


]
\input epsf

%68.35.Bs - Surface structure and topography
%68.35.Ja - Surface and interface dynamics and vibrations
%68.55.-a - Thin film structure and morphology
%68.55.Jk - Structure and morphology; thickness

\newpage

\section{Introduction}

The morphological evolution of crystalline surfaces has long been a
major
focus of attention in surface science. At low temperatures, below the
roughening temperature of a low index facet plane, this is often
dominated
by the motion of surface steps, typically produced by miscuts or arising
from dislocations \cite{williamsreviews}. Vicinal surfaces, created by a
small miscut to the facet plane, have steps all of the same sign (either
up
or down steps). They offer a particularly simple testing ground for the
study
of step models of kinetic processes and their connections to physics
both on
atomic and macroscopic scales.

In this paper, we study the kinetics of faceting and relaxation of
vicinal
surfaces using one-dimensional (1D) models of straight steps. We show
that
on long length and time scales the surface profiles exhibit a scaling
behavior in many cases. This general conclusion is not at all
surprising; it
agrees with the classic work by Mullins \cite{mullins}, who first
investigated problems of this kind using a simple continuum model.
However,
the details of the calculations we carried out and the nature of the
scaling
functions describing the surface profiles are rather different. These
differences arise because we take the continuum limit of a physical
model
where steps play the fundamental role in the kinetic processes
describing
surface evolution. Even in the continuum limit there are features of the
scaled profiles that reflect this underlying physics. There has been
much
discussion in the literature about the derivation of continuum equations
from discrete models \cite{williamsreviews}. Our work shows in detail
how
this can be done in some simple cases and illustrates some of the subtle
issues that can arise.

Mullins \cite{mullins} used a simple 1D continuum model to describe the
growth of a single linear facet on a vicinal surface, assuming that the
surface free energy of the complex surface away from the facet is an
analytic function of the local surface slope. This assumption is valid
in
many cases since the vicinal surface itself is rough
\cite{williamsreviews}.
However, since the existence of a facet is associated with the breakdown
of
analyticity of the surface free energy as a function of slope, this
approach
will fail whenever the local slope approaches zero, the slope of the
flat
facet. There is nothing in the usual continuum approach to prevent this
from
happening. Indeed, as noted below, in some cases the standard continuum
model predicts that surface profiles will oscillate sufficiently about
the
slope of the vicinal surface to produce regions with {\em negative}
local
slope. This implies the creation of new {\em anti-steps} (steps of the
opposite sign), an energetically costly process that does not occur in
the
step models we consider, even in the continuum limit. Furthermore, since
at
low temperatures steps play an essential role in determining surface
kinetic
processes, the effective kinetic coefficients in a continuum model must
depend on the step density in ways that may seem hard to understand when
viewing the system from a continuum perspective from the outset.

To examine these issues in detail, we start with a 1D model of discrete
straight steps and describe surface morphological changes in terms of
their
motion. The equations of motion for the individual steps reflect the
mass
transport mode and are derived in the standard way
\cite{williamsreviews},
using a linear kinetics assumption based on the difference in the
chemical
potentials for each step arising from step repulsions. By considering
length
scales large compared to the step spacing, we show that it is possible
to
take the continuum limit of these equations in a consistent way. The
dynamic
equation for the evolution of the local slope, or the {\em step density
function}, is thus obtained systematically from the equations for
individual
step motion. The long time evolution of this dynamic equation is then
investigated with a scaling ansatz. We applied this scheme to three
different 1D physical systems: reconstruction driven faceting,
relaxation of
an infinite bunch, and flattening of a groove.
In all cases, the scaling function is described by the same differential
equation for the same mass transport mode; the solutions differ only
because
of different boundary conditions for the scaling functions. As a result,
the
values of the scaling exponents depend only on the mass transport
mechanism,
and are consistent with the predictions of Mullins' classical theory.
However, the scaling functions themselves --- the scaled slopes of the
surface profile --- differ from Mullins' results and are in excellent
agreement with numerical solutions of the discrete equations.

The paper is organized as follows. We introduce the 1D step models in
Sec.
II. In Sec. III the step density function is first defined. Then we
introduce a scaling ansatz and derive a differential equation for the
scaling functions. In Sec. IV, the properties of the scaling function
are
further investigated. In Sec. V, we check the validity of the scaling
analysis using three different step flow models. The shape of the
scaling
function
is obtained by numerical integration of the differential equation. We
also
show that this scaling function coincides with the result of the
simulations
of the discrete step flow models. Our conclusions are given in Sec. VI.

\section{One Dimensional Step Flow Models}

Below the roughening temperature of a high symmetry orientation of the
crystal, a vicinal surface consists of flat terraces separated by atomic
steps. Ignoring islands and vacancies, the morphological evolution of
the surface is a consequence of exchange of atoms between steps and
their neighboring terraces resulting in motion of the steps. An analysis
of surface evolution in terms of step flow was first carried out by
Burton, Cabrera and Frank\cite{BCF} and this has been
generalized by many other authors\cite{williamsreviews}. In what 
follows we describe the evolution of vicinal surfaces using a 
simple step flow model.

We consider two
limiting channels for mass transport involving the terrace adatoms. In
the
first case the adatom mass flow on each terrace is {\em local, }and{\em
\ }
takes place{\em \ }by surface diffusion. In the second case the terrace
adatoms can easily exchange with a {\em global} reservoir, perhaps
through
direct hops to distant regions of the surface or by rapid exchange with
the
vapor. We refer to these limiting cases as the local (LEM) and global
(GEM)
exchange mechanisms, respectively. They correspond to the surface
diffusion
and evaporation-condensation mass transport mechanisms considered by
Mullins 
\cite{mullins}.

\subsection{Local Mass Exchange Mechanism}

Consider an array of flat terraces separated by straight parallel steps
with
horizontal positions, $x_{n}$. The index $n$
grows in the direction of positive surface slope. These steps may absorb
or
emit atoms which then diffuse across the neighboring terraces. We ignore
evaporation. Assuming attachment-detachment limited kinetics (i.e., that
diffusion on terraces is very fast compared to the rate of attachment
and
detachment of atoms to and from step edges), the adatoms which diffuse
on
the $n$th terrace maintain a {\em uniform} chemical potential
$\mu_{n}^{t}$
across the terrace\cite{comment_for_the_general_case}. 

We further assume that the flux of atoms at the two step edges bounding
the $n$th terrace is determined by first order kinetics, characterized
by a
(for
simplicity symmetric) attachment-detachment rate coefficient $k$: 
\begin{eqnarray}
J_{n}^{+} &=&k\left( \mu _{n}^{t}-\mu _{n}^{s}\right) \;,  \nonumber \\
J_{n+1}^{-} &=&k\left( \mu _{n}^{t}-\mu _{n+1}^{s}\right) \;.
\label{eq:boundary}
\end{eqnarray}
Here $J_{n}^{+}$ and $J_{n}^{-}$ denote the flux from the lower and
upper
neighboring terraces into the $n$th step respectively. $\mu _{n}^{s}$ is
the
step chemical potential associated with adding an adatom to the $n$th
step.
In the case of elastic \cite{MarchenkoParshin,AndreevKosevich} or
entropic
repulsive interactions between steps, it is well known that the step
chemical potential then takes the form
\cite{OzdemirZangwill,RettoriVillain} 
\begin{equation}
\mu _{n}^{s}=2g\left[ \frac{1}{(x_{n}-x_{n-1})^{3}}-
\frac{1}{(x_{n+1}-x_{n})^{3}}\right] \;,  
\label{eq:chemical_potential}
\end{equation}
where $g$ is the strength of the repulsive interactions.

Next, we assume that diffusion processes are fast compared with the
motion
of steps. Within this quasi-static approximation, the density of adatoms
on
the terraces reaches a steady state for each step configuration. In this
steady state we have $J_{n}^{+}+J_{n+1}^{-}=0$ for any $n$, and
therefore
the free energy associated with the addition of an adatom on the $n$th
terrace, $\mu _{n}^{t}$, takes the form 
\begin{equation}
\mu _{n}^{t}=\frac{\mu _{n}^{s}+\mu _{n+1}^{s}}{2}\;.  \label{eq:Cns}
\end{equation}
Combining mass conservation at the $n$th step with Eqs.\
(\ref{eq:boundary})
and (\ref{eq:Cns}), we obtain the following expression for the step
velocity: 
\begin{equation}
\dot{x}_{n}=-a^{2}\left( J_{n}^{+}+J_{n}^{-}\right)
=\frac{a^{2}k}{2}\left(
2\mu _{n}^{s}-\mu _{n+1}^{s}-\mu _{n-1}^{s}\right) \;,
\label{eq:lem_velocity}
\end{equation}
with $a$ denoting the lattice constant of the crystal.

\subsection{Global Mass Exchange Mechanism}

As with the LEM, we assume linear kinetics; i.e., the flux of atoms from
the
reservoir to the $n$th step is proportional to the difference between
chemical potentials of the reservoir and the step: 
\begin{equation}
J_n=k\left(\mu_{res}-\mu^s_n\right)\;.
\end{equation}
Here $\mu_{res}$ is the reservoir chemical potential, which we choose as
the
chemical potential of a flat surface; i.e., $\mu_{res}=0$. The velocity
of
the $n$th step in this case is simply 
\begin{equation}
\dot{x}_n=-a^2J_n=a^2k\mu^s_n\;.  \label{eq:gem_velocity}
\end{equation}
Although the exchange rates in the LEM and GEM cases are different we
use
the symbol $k$ to denote both of them. It should be clear from the
context
which exchange rate we refer to.

\section{Scaling Analysis and Continuum Models}
Simulations of the LEM and GEM step flow models \cite{JeongWeeks2}
suggest
that the behavior of these systems at long length and time scales can be
described in terms of a {\em step density function} $D$ (i.e., the
inverse
step separation) that is a continuous function of both position and
time. Generally speaking, continuum descriptions of step systems should
be 
valid when every typical surface feature consists of very many 
steps. In the cases we study here, the simulations indicate that the
step
density function scales in 
time according to 
\begin{equation}
D(x,t)=F(xt^{-1/\gamma })\;,  \label{eq:scaling_ansatz}
\end{equation}
with a positive exponent $\gamma$. Thus, the number of steps in every
surface feature grows with time as $t^{1/\gamma}$.
We should therefore be able to accurately describe the 
evolution of the system in terms of a continuum model in the long 
time limit. 

Eq.\ (\ref{eq:scaling_ansatz}) makes the even stronger
assertion that the  
surface features have a self-similar shape during their evolution, 
and can be related by a proper rescaling of time and distance.
This scenario is similar to the scaling
exhibited by a decaying crystalline cone, which two of us have studied
\cite{cone-short,cone-long}. 
In this section we carry out a scaling analysis,
similar to the one in Refs.\ \onlinecite{cone-short} and
\onlinecite{cone-long}, to obtain the
scaling
exponents $\gamma $ and the differential equation for the scaling
function $F
$. We also study the effects of the different mass exchange mechanisms.

We start by defining the step density function in the middle of the
terraces: 
\begin{equation}
D\left( \frac{x_{n}+x_{n+1}}{2},t\right) \equiv
\frac{1}{x_{n+1}-x_{n}}\;.
\label{eq:density_def}
\end{equation}
Assuming continuity, the full time derivative of the step density is
given
by: 
\begin{equation}
\frac{dD}{dt}=\frac{\partial D}{\partial t}+\frac{\partial D}{\partial
x}
\cdot \frac{dx}{dt}\;.  \label{eq:full_t_deriv}
\end{equation}
Eq.\ (\ref{eq:full_t_deriv}) can be evaluated in the middle of the
terrace
between two steps (i.e., at $x=(x_{n}+x_{n+1})/2$). 
Assuming that the scaling ansatz Eq.\ (\ref{eq:scaling_ansatz}) holds,
we now
change variables to $\theta \equiv t^{1/\gamma }$ and $\xi _{n}\equiv
x_{n}\theta ^{-1}$, and transform Eq.\ (\ref{eq:full_t_deriv}) into an
equation
for the scaling function $F$ evaluated at $\xi=\left(\xi _{n}+\xi
_{n+1}\right)/2$:
\begin{equation}
\frac{dF}{d\xi }\left( \theta ^{\gamma
-1}\frac{\dot{x}_{n+1}+\dot{x}_{n}}{2}-\frac{\xi }{\gamma }\right) 
+F^{2}\theta ^{\gamma }\left(\dot{x}_{n+1}-\dot{x}_{n}\right) =0\;,  
\label{eq:partial2}
\end{equation}
In going from Eq.\ (\ref{eq:full_t_deriv}) to Eq.\
(\ref{eq:partial2}) we have used the fact that
$dD/dt=-D^{2}(\dot{x}_{n+1}-\dot{x}_{n})$ and that
$dx/dt=(\dot{x}_{n}+\dot{x}_{n+1})/2$.
$\dot{x}_{n}$, $\dot{x}_{n+1}$ themselves are now expressed in term of
the
$\xi_{n}$'s using Eq.\ (\ref{eq:lem_velocity}) or
(\ref{eq:gem_velocity}) depending on the exchange mechanism.

Let us also rewrite Eq.\
(\ref
{eq:density_def}) in terms of $\theta ,$ $F,$ and the $\xi _{n}$'s: 
\begin{equation}
\xi _{n+1}-\xi _{n}=\frac{\theta ^{-1}}{F\left[ (\xi _{n+1}+\xi
_{n})/2\right] }\;.  \label{eq:xisdifference}
\end{equation}
According to this, the difference between successive $\xi _{n}$'s is of
order $\theta ^{-1}$ wherever $F$ does not vanish. In the large $\theta
$
(long time) limit these differences become vanishingly small. The
differences $\xi _{n+k}-\xi $ will also be small as long as $k$ is
finite.
We can therefore take the continuum limit in the variable $\xi $,
consistent
with our original supposition.

To this end, we evaluate the function $F$ at the position $(\xi
_{n+k}+\xi
_{n+k+1})/2$ by using its Taylor expansion 
\begin{eqnarray}
&F&\left( \frac{\xi _{n+k}+\xi _{n+k+1}}{2}\right) \equiv \frac{\theta
^{-1}}{
\xi _{n+k+1}-\xi _{n+k}} \nonumber \\
&=&\sum_{m=0}^{\infty
}\frac{1}{m!}\frac{d^{m}F(\xi )}{
d\xi ^{m}}\left( \frac{\xi _{n+k}+\xi _{n+k+1}}{2}-\xi \right) ^{m}\;.
\label{eq:taylorexpansion}
\end{eqnarray}
Next, we expand 
\begin{equation}
\xi _{n+k}=\xi +\sum_{m=1}^{\infty }\phi _{km}\theta ^{-m}\;,
\label{eq:xisexpansion}
\end{equation}
and insert this into Eq.\ (\ref{eq:taylorexpansion}). By equating terms
of
the same order in $\theta ^{-1}$ on both sides of Eq.\ (\ref
{eq:taylorexpansion}), we can find the coefficients $\phi _{km}$ for any
desired values of $k$ and $m$. These coefficients involve the function
$F$
and its derivatives evaluated at $\xi \equiv (\xi _{n}+\xi _{n+1})/2$.

Having found the expansion coefficients $\phi_{km}$ we now return to
Eq.\ (\ref{eq:partial2}). 
This equation depends on the velocities $\dot{x}_n$
and $\dot{x}_{n+1}$ which in turn depend on 
$\xi_{n-2}$ ... $\xi_{n+3}$ or $\xi_{n-1}$ ... $\xi_{n+2}$ in the
 LEM or GEM cases respectively. Using
Eq.\
(\ref{eq:xisexpansion}) we expand Eq.\ (\ref{eq:partial2}) in the LEM
case with the
following result:
\begin{equation}  \label{eq:local_expand_diff}
\theta^{\gamma-4}a^2kg\frac{d^2}{d\xi^2}\left(\frac{1}{F}\frac
{d^2}{d\xi^2}\frac{3F^2}{2}\right)
-\frac{\xi}{\gamma}\frac{dF}{d\xi}+{\cal
O}\left(\theta^{\gamma-6}\right)=0\;.
\end{equation}

Consider Eq.\ (\ref{eq:local_expand_diff}). It involves different powers
of
the scaled time $\theta$ and cannot be satisfied at all times (unless
$F$ is
trivially independent of $\xi$). However in the $\theta \rightarrow
\infty$
(long time) limit Eq.\ (\ref{eq:local_expand_diff}) can be satisfied
exactly. This can be achieved by setting the value of the scaling
exponent
to be $\gamma=4$ and then requiring the first and second terms to cancel
each other. Neglecting the small third term we are left
with
the long time limit of the LEM differential equation 
\begin{equation}
4a^2kg\frac{d^2}{d\xi^2}\left(\frac{1}{F}\frac{d^2}{d\xi^2}\frac{3F^2}{2
}\right) -\xi\frac{dF}{d\xi}=0\;.  
\label{eq:local_diff}
\end{equation}
This equation determines the scaling function $F$ and is exact only in
the
long time limit. In this limit it is equivalent to the
attachment/detachment limited scaling
equation suggested by Liu et
al\cite{DaJiangFuJohnsonWeeksWilliams}. It also agrees with 
Nozi\`{e}res' continuum treatment\cite{Nozieres} of
attachment/detachment
limited kinetics provided one assumes scaling.
Thus, in the long time limit, our analysis confirms the validity of
previous continuum
treatments.  
However at any finite time there are corrections of order
$\theta
^{-2}$ to the scaling solution $D(x,t)=F(\xi )$. 

A similar procedure can be applied in the GEM case. It leads to the
scaling
exponent $\gamma=2$ and the following equation for the scaling function: 
\begin{equation}
4a^2kg\frac{d^2 F^3}{d\xi^2}+\xi\frac{dF}{d\xi}=0\;.
\label{eq:global_diff}
\end{equation}
As in the LEM case, the leading correction to the density function
decays as 
$\theta ^{-2}$. Like Mullins, we obtain a fourth order equation with LEM
and
a second order equation with GEM. However, these equations have a more
complicated form than those arising from Mullins' continuum model and
will
have different solutions.

To conclude this part let us emphasize a few points regarding the above
analysis. Eqs. (\ref{eq:local_diff}) and (\ref{eq:global_diff}) together
with the scaling ansatz Eq.\ (\ref{eq:scaling_ansatz}) and the values of
$\gamma $
constitute a continuum model for the surface dynamics. This model was
derived directly from the discrete step system and is {\em exact} in the
long time limit. In addition, the above scaling analysis is robust in
the
following sense. The values of the scaling exponent $\gamma $ are not
sensitive to the exact nature of the step-step interactions in the
discrete
model. In Appendix A we show that for a general interaction $\gamma=4$
and $\gamma=2$ in the LEM and GEM cases respectively. However, the
differential
equations for the scaling functions do depend on the form of the
interaction.

\section{Properties Of The Scaling Function}

Here we study some properties of our continuum model. Our purpose is to
derive several relations for the scaling function, which hold in
general.
These relations are useful in the derivation of the boundary conditions
necessary in order to solve Eqs.\ (\ref{eq:local_diff}) and (\ref
{eq:global_diff}) for the scaling functions of various systems.

First, note that the scaling function must have a finite limit,
$F_{\infty}$, at infinity. 
Otherwise the step density there would change infinitely
fast. We choose the unit of length so that $F_{\infty}=1$. We also
choose
the unit of time so that $4a^2kg=1$ in Eqs.\ (\ref{eq:local_diff}) and
(\ref
{eq:global_diff}).

Next, we investigate the time dependence of the volume and the total
number
of steps in the system. It turns out that these quantities can be
calculated
directly from the differential equations. 
The change in the volume of the system in the positive $x$ half of space
during the time interval from zero to $t$ is given by 
\begin{equation}
\Delta V=\int_0^\infty\left[h(x,t)-h(x,0)\right]dx\;.
\end{equation}
$h(x,t)=\int_0^xD(\tilde{x},t)d\tilde{x}$ is the profile height measured
in units of the lattice constant.

Integrating by parts and changing to scaling variables we obtain the
equation
\begin{equation}
\Delta V=\theta^2\int_0^\infty\left(
F_\infty-F\right)\xi d\xi\;,
\end{equation}
where we have ignored the surface term assuming no evolution
occurs infinitely far from the origin.

Let us calculate the last integral in the LEM case. According to Eq.
(\ref
{eq:local_diff}) 
\begin{equation}
\int_{\xi_a}^{\xi_b}\xi\frac{d^2}{d\xi^2}\left(\frac{1}{F}\frac{d^2}
{d\xi^2}\frac{3F^2}{2}\right)d\xi=\int_{\xi_a}^{\xi_b} \xi^2
\frac{dF}{d\xi}d\xi\;,
\end{equation}
where $\xi_a$ and $\xi_b$ are boundaries of a region where Eq. (\ref
{eq:local_diff}) is valid. Integrating each term by parts we find that 
\begin{eqnarray}
\Delta &V&_{LEM}=\nonumber \\
&\frac{\theta^2}{2}&\left. \left[ \xi^2\left(F_\infty-F
\right)+\xi\frac{d}{d\xi}\left(\frac{1}{F}\frac{d^2}{d\xi^2}\frac{3F^2}{
2}\right)-\frac{1}{F}\frac{d^2}{d\xi^2}\frac{3F^2}{2}\right]
\right|_{\xi_a}^{\xi_b}  \nonumber \\
&+&\theta^2\int_{\xi \notin (\xi_a,\xi_b)}\left(F_\infty-F \right)\xi
d\xi
\;.  \label{eq:LEM_vchange}
\end{eqnarray}
Similarly, we can calculate the volume change in the GEM case using Eq.
(\ref
{eq:global_diff}) and obtain the equation 
\begin{eqnarray}
\Delta V_{GEM}=&\frac{\theta^2}{2}&\left.\left[\xi^2\left(F_\infty-F
\right)-\xi\frac{dF^3}{d\xi}+F^3\right]\right|_{\xi_a}^{\xi_b}
\nonumber \\
+&\theta^2&\int_{\xi \notin (\xi_a,\xi_b)}\left(F_\infty-F \right)\xi
d\xi
\;.  \label{eq:GEM_vchange}
\end{eqnarray}

A similar treatment can be applied to calculate changes in the number of
steps in the positive $x$ half of the system, and we find 
\begin{equation}
\Delta N=\theta \int_0^\infty\left(F-F_\infty \right)d\xi\;.
\end{equation}
Again we can evaluate this integral by using Eqs. (\ref{eq:local_diff})
and
( \ref{eq:global_diff}). The results are 
\begin{eqnarray}
\Delta N_{LEM}=&\theta&\left.\left[\xi\left(F-F_\infty
\right)-\frac{d}{d\xi}\left(\frac{1}{F}\frac{d^2}{d\xi^2}\frac{3F^2}{2} 
\right) \right]
\right|_{\xi_a}^{\xi_b}  \nonumber \\
+&\theta&\int_{\xi \notin (\xi_a,\xi_b)}\left(F-F_\infty \right) d\xi \;
\label{eq:LEM_nchange}
\end{eqnarray}
in the LEM case and 
\begin{eqnarray}
\Delta N_{GEM}=&\theta&\left.\left[ \xi\left(F-F_\infty \right)
+\frac{dF^3}{d\xi}\right] \right|_{\xi_a}^{\xi_b}  \nonumber \\
+&\theta&\int_{\xi \notin (\xi_a,\xi_b)}\left(F-F_\infty \right) d\xi \;
\label{eq:GEM_nchange}
\end{eqnarray}
in the GEM case.

Finally we study the behavior of the scaling functions near regions of
zero
step density. We recall that our scaling analysis is valid only in
regions
of space where the step density does not vanish (see Eq. (\ref
{eq:xisdifference})). Points of vanishing step density should therefore
be
treated separately. Assume that $\xi _{0}$ is such a point for which
$F(\xi
_{0})=0$. Expanding the scaling function in powers of $\xi -\xi _{0}$ in
the
vicinity of $\xi _{0}$ and using Eqs.\ (\ref{eq:local_diff}) and (\ref
{eq:global_diff}), we find that in the LEM case 
\begin{equation}
F=\sum_{n=1}^{\infty }b_{n}(\xi -\xi _{0})^{n/2}\;,
\label{eq:small_F_exp_lem}
\end{equation}
and in the GEM case 
\begin{equation}
F=\sum_{n=1}^{\infty }b_{n}(\xi -\xi _{0})^{n/3}\;.
\label{eq:small_F_exp_gem}
\end{equation}
Thus a point of vanishing step density is a {\em singular point} of the
scaling function at which all its derivatives diverge.

\section{Examples}

To check the validity of our scaling analysis, we consider three
different
step flow models, which according to numerical simulations obey scaling
under both exchange mechanisms. All three systems consist of straight
parallel and initially equidistant steps, but differ in their boundary
conditions.

\subsection{Reconstruction Driven Faceting}

\begin{figure}[h]
\epsfxsize=50mm
\centerline{\epsffile{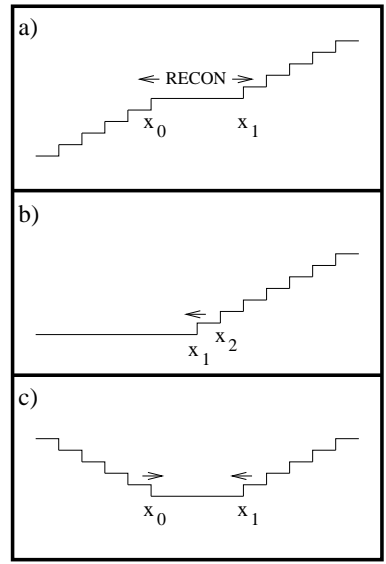}}
\vspace{3mm}
\caption{The three systems considered in this section: (a)
reconstruction
driven faceting, (b) relaxation of an infinite bunch and (c) flattening
of a
groove.}
\label{example_systems}
\end{figure}

The first example is that of reconstruction driven faceting studied by
Jeong
and Weeks in Refs.\ \onlinecite{JeongWeeks2} and
\onlinecite{JeongWeeks1}. It is mathematically
equivalent to the model of facet growth during thermal etching studied
by
Mullins \cite{mullins}. Within their model, surface reconstruction that
lowers the free energy can nucleate only on terraces of width larger
than a
critical width, $w_{c}$. Therefore, terraces of width $w>w_{c}$ have a
lower
free energy than those of width $w<w_{c}$. We consider the evolution of
such
a step system starting from a configuration where all the terraces
except
one have the same width $w<w_{c}$. (For simplicity, we do not permit
reconstruction on other terraces, even if their widths exceeds $w_{c}$
during the surface evolution. The possibility of such ``induced
nucleation''
on other terraces is discussed in Refs.\ \onlinecite{JeongWeeks2} and
\onlinecite{JeongWeeks1}.)
The
zeroth terrace (between $x_{0}$ and $x_{1}$ in Fig.\
\ref{example_systems}
(a)) is of different width, larger than $w_{c}$ and is reconstructed.
This
reconstructed terrace tends to become even wider, and this is reflected
as a
shift of magnitude $\epsilon $ in the chemical potentials of the two
steps: 
\begin{eqnarray}
\mu _{0}^{s} &=&2g\left[ \frac{1}{(x_{0}-x_{-1})^{3}}-
\frac{1}{(x_{1}-x_{0})^{3}}-\epsilon \right] \;,  \nonumber \\
\mu _{1}^{s} &=&2g\left[ \frac{1}{(x_{1}-x_{0})^{3}}-
\frac{1}{(x_{2}-x_{1})^{3}}+\epsilon \right] \;.  
\label{eq:RDchemicalpotentials}
\end{eqnarray}
As a result, the step at $x_{0}$ propagates to the left, while the one
at $x_{1}$ propagates to the right. In the long time limit the step
density
of the system obeys scaling. In particular, the size of the facet at the
origin grows as $t^{1/\gamma }$ in both LEM and GEM cases, with $\gamma
=4$
in the LEM case and $\gamma=2$ in the GEM case.

In order to compare our scaling analysis to simulation results we have
to
solve Eqs.\ (\ref{eq:local_diff}) and (\ref{eq:global_diff}) with the
relevant boundary conditions. Since the system is symmetric about the
origin
it is sufficient to solve the scaling function $F$ for positive $\xi$.
In
addition, our expansion in the small parameter $\theta^{-1}$ is valid
only
in regions where $F$ does not vanish (see Eq.\
(\ref{eq:xisdifference})).
This requirement is violated on the diverging facet around the origin.
Therefore, $F$ obeys Eqs.\ (\ref{eq:local_diff}) or
(\ref{eq:global_diff})
only for $\xi \geq \xi_1$, where $\xi_1$ is the scaled position of the
first
step, and $F(\xi)=0$ for $\xi<\xi_1$.

\subsubsection{Local Mass Exchange Mechanism}

The first boundary condition is set by our choice of the value of $F$ at
infinity, namely $F_\infty=1$.

Next, consider the number of steps, which is a conserved quantity in our
step model. This is an important difference from the continuum approach
of
Mullins. Using Eq. (\ref{eq:LEM_nchange}) with $\xi _{a}=\xi _{1}$, $\xi
_{b}=\infty $ and the condition $F(\xi )=0$ for $\xi <\xi _{1}$, we find
that in the LEM case 
\begin{equation}
F'''(\xi _{1})=\left. \left( \frac{F'^3}{
F^{2}}-
\frac{2F'F''}{F}+\frac{\xi F}{3}\right) \right|
_{\xi
_{1}}\;,  \label{eq:RDF_LEM_bc2}
\end{equation}
where primes denote derivatives with respect to $\xi$.

Two additional boundary conditions can be found by investigating the
velocity of the first step. According to our scaling ansatz the position
of
the first step is $x_1=\theta \xi_1$. In the long time scaling limit the
velocity of this step goes to zero as $\theta^{-3}$. Using Eq. (\ref
{eq:lem_velocity}) together with the symmetry of the system we obtain
the
following expression for $\dot{x}_1$: 
\begin{equation}
\dot{x}_1=\frac{1}{4}\left[\frac{3}{(2x_1)^3}-\frac{4}{(x_2-x_1)^3}+
\frac{1 
}{(x_3-x_2)^3}+3\epsilon\right]\;.  \label{eq:RDF_LEM_v1}
\end{equation}

Rewriting Eq. (\ref{eq:RDF_LEM_v1}) in terms of the scaling variables
and
expanding in $\theta ^{-1}$ we find that 
\begin{equation}
\dot{x}_{1}=\frac{3}{4}\left( \epsilon -F^{3}\right) +\frac{3FF'\theta
^{-1}}{4}+{\cal O}\left( \theta ^{-2}\right) \;,
\label{eq:x1dot_expansion}
\end{equation}
where $F$ is evaluated at $(\xi _{1}+\xi _{2})/2$. In order for
$\dot{x}_{1}$
to vanish as $\theta ^{-3}$ we must have $F[(\xi _{1}+\xi
_{2})/2]=\epsilon
^{1/3}$ and $F'[(\xi _{1}+\xi _{2})/2]=0$. Terms of order
$\theta
^{-2}$ on the r.h.s.\ also have to vanish, but they include corrections
to
scaling, which we ignore in this work. In the long time limit the
difference
between $\xi _{1}$ and $(\xi _{1}+\xi _{2})/2$ is negligible and we
arrive
at the boundary conditions 
\begin{equation}
F(\xi _{1})=\epsilon ^{1/3},\;\mbox{and}\;\;\;F'(\xi _{1})=0\;.
\label{eq:RDF_LEM_bc3&4}
\end{equation}

We solved Eq.\ (\ref{eq:local_diff}) numerically by applying the three
boundary conditions at $\xi =\xi _{1}$ (Eqs.\ (\ref{eq:RDF_LEM_bc2}) and
(\ref{eq:RDF_LEM_bc3&4})), and by tuning $F''(\xi _{1})$
and $\xi _{1}$ itself to satisfy the boundary condition at
infinity. In the upper half of Fig.\ \ref{rdf_scaling} we compare the
resulting
solution
with scaled density functions taken from numerical simulations of the
discrete model. The agreement is quite impressive. These results should
be
compared with Fig.\ 8 in Ref.\ \onlinecite{mullins}, which predicts
anti-step
formation
for large values of the slope parameters $m/n$.

\subsubsection{Global Mass Exchange Mechanism}
Like in the LEM case, the requirement that the velocity of the first
step decays in time results in the boundary condition
\begin{equation}
F(\xi _{1})=\epsilon ^{1/3}\;.  \label{eq:RDF_GEM_bc1}
\end{equation}
Again $\xi_1$ is the scaled position of the first step.

Another boundary condition is derived from the conservation of the
number of
steps. Eq.\ (\ref{eq:GEM_nchange}) with $\xi_a=\xi_1$, $\xi_b=\infty$
and
the condition $F(\xi)=0$ for $\xi<\xi_1$ implies that 
\begin{equation}
F'(\xi_1)=-\frac{\xi_1}{3F(\xi_1)}\;.  \label{eq:RDF_GEM_bc2}
\end{equation}
These two conditions are sufficient for solving Eq.\
(\ref{eq:global_diff})
when the value of $\xi_1$ is known.

\begin{figure}[h]
\epsfxsize=80mm
\centerline{\epsffile{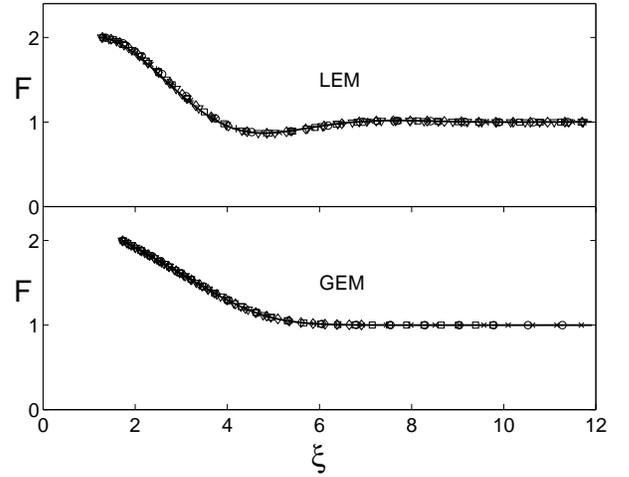}}
\vspace{3mm}
\caption{Solution of the reconstruction driven faceting scaling
function in the LEM and GEM cases
(solid lines)
compared with scaled density functions from numerical simulations with
$\epsilon=8$. Different symbols represent density functions at
different times.}
\label{rdf_scaling}
\end{figure}

Finally we tune $\xi_1$ to satisfy the boundary condition at
infinity. The resulting 
solution and its comparison with simulation data are shown in the
lower half of Fig.\ \ref{rdf_scaling}.

\subsection{Relaxation of an Infinite Bunch}

In the second example we study the relaxation of an infinite bunch of
steps.
The initial step configuration (Fig.\ \ref{example_systems} (b))
consists of
an infinite facet at $x<0$, in contact with an infinite array of
uniformly
spaced steps (at $x>0$). The first step has a single neighbor, and
therefore
its chemical potential is 
\begin{equation}
\mu^s_1=\frac{-2g}{(x_2-x_1)^3}\;.  \label{eq:IBchemicalpotential}
\end{equation}
In the LEM case there is a complication, since the first step may
exchange
adatoms with the infinite facet on its left. Hence we have to 
specify $\mu^t_0$, the adatom chemical potential on the facet. We assume
here
that $\mu^t_0=\mu^s_1$, neglecting any exchange of atoms between the
first
step
and the facet. As a result the volume is conserved in the LEM case.

Simulations of this system in both the LEM and GEM cases suggest that
the
system exhibit scaling with the origin of the scaled position at the
initial
position of the first step. The leftmost steps from
the
bunch move to the left into the facet due to the repulsive interactions.
The
first step recedes in time and its position scales as $x_{1}\sim
-t^{1/\gamma }$. At the same time the separation between the first steps
grows. For $\xi <\xi _{1}$, the scaled position of the first step, the
step
density always vanishes, and we have to find the scaling function only
for $\xi _{1}\leq \xi <\infty $.

\subsubsection{Local Mass Exchange Mechanism}

The velocity of the first step vanishes in the scaling limit. We can
therefore evaluate $F(\xi _{1})$ by expanding $\dot{x}_{1}$
in $\theta ^{-1}$, and requiring that the zeroth order term vanish. We
find
that the zeroth order term in $\dot{x}_{1}$ is proportional to
$F^{3}(\xi
_{1})$, which implies that $F(\xi _{1})=0$.

Two additional boundary conditions can be derived from the conservation
of
volume and the number of steps. Using Eqs.\ (\ref{eq:LEM_vchange}) and
(\ref
{eq:LEM_nchange}) to calculate changes in the volume and number of steps
in
the positive part of the $\xi$ axis together with equivalent equations
for $\xi<0$, we find that 
\begin{equation}
\left.
\left[\xi\frac{d}{d\xi}\left(\frac{1}{F}\frac{d^2}{d\xi^2}\frac{3F^2}{2}
\right)-\frac{1}{F}\frac{d^2}{d\xi^2}\frac{3F^2}{2}\right]\right|_{\xi_1
}=0\;,  \label{eq:inf_bunch_lem_bc3}
\end{equation}
and 
\begin{equation}
\left.\frac{d}{d\xi}\left(\frac{1}{F}\frac{d^2}{d\xi^2}\frac{3F^2}{2}
\right) \right|_{\xi_1}=0\;.  \label{eq:inf_bunch_lem_bc4}
\end{equation}
As mentioned above, a zero of the scaling function is a singular point
at
which all derivatives of $F$ diverge. Since $F(\xi_1)=0$, the limit 
$\xi\rightarrow \xi_1$ in the last two boundary conditions should be
taken
with care. This can be done by considering the power series (\ref
{eq:small_F_exp_lem}). The differential equation (\ref{eq:local_diff})
and
the boundary conditions $F(\xi_1)=0$, (\ref{eq:inf_bunch_lem_bc3}) and
(\ref
{eq:inf_bunch_lem_bc4}) impose connections between the coefficients of
the
expansion and leave only $b_1$ as a free parameter.

We now use the following procedure to compute $F$. For given values of
$b_1$
and $\xi_1$ we approximate $F$ at $\xi_1+\delta \xi$ using the series
expansion (\ref{eq:small_F_exp_lem}) with suitable truncation. 
At $\xi_1+\delta \xi$ the derivatives of $F$ are finite, and we solve
$F$
from
there by numerical integration. We adjust the values of $b_1$ and
$\xi_1$ in
order to satisfy the boundary condition at infinity,
thus
obtaining the scaling function $F(\xi)$ in the LEM case. In the upper
half of Fig.\ \ref
{inf_bunch_scaling} we compare this solution with simulation data.

\subsubsection{Global Mass Exchange Mechanism}

\begin{figure}[h]
\epsfxsize=80mm
\centerline{\epsffile{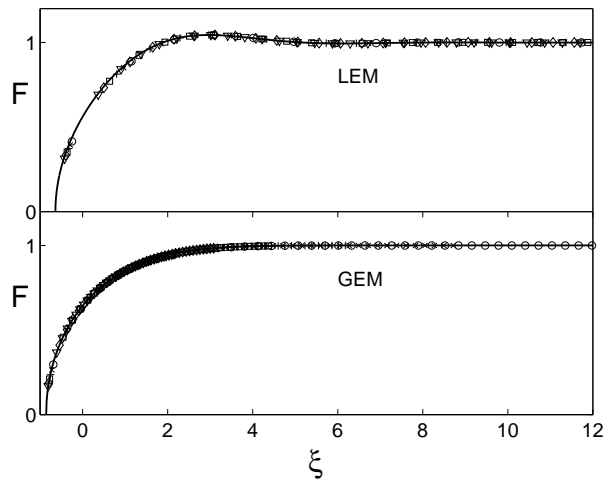}}
\vspace{3mm}
\caption{Solution of the infinite bunch scaling function
in the LEM and GEM cases  (solid lines)
compared with scaled density functions from numerical simulations of
the discrete model. Different symbols represent density functions at
different
times.} 
\label{inf_bunch_scaling}
\end{figure}

Since $\dot{x}_{1}\propto -(x_{2}-x_{1})^{-3}$, the requirement that the
velocity of the first step vanishes in the scaling limit leads to the
boundary condition 
\begin{equation}
F(\xi _{1})=0\;.
\end{equation}
To derive another boundary condition we use Eq.\ (\ref{eq:GEM_nchange})
and
impose conservation of the number of steps. The following relation is
thus
obtained: 
\begin{equation}
\left. \frac{dF^3}{d\xi}\right| _{\xi _{1}}=0\;.
\end{equation}
This implies that all the coefficients in the series expansion (\ref
{eq:small_F_exp_gem}) diverge except for $b_{1}$ which vanishes. Thus
the
convergence radius of (\ref{eq:small_F_exp_gem}) is zero and it is
difficult
to calculate the scaling function with the numerical procedure used in
the
LEM case. Nevertheless, taking a small enough value of $b_{1}$ and
tuning $\xi _{1}$ to satisfy the boundary condition at infinity, 
we were able to
calculate $F$ approximately. In the lower half of Fig.\
\ref{inf_bunch_scaling} we compare
this approximation with simulation data.

\subsection{Flattening Of A Groove}

Our last example is the flattening of a groove cut in the crystal
surface.
The initial configuration (Fig.\ \ref{example_systems} (c)) consists of
two
infinite step bunches with steps of opposite signs that meet at the
origin.
Step repulsion within each bunch pushes the bottom step and anti-step
towards each other until they collide and annihilate. We assume here
that
steps of opposite sign do not interact. Thus the chemical potential of
the
two bottom steps includes interaction with only one neighboring step,
namely 
\begin{equation}
\mu^s_1=\frac{-2g}{(x_2-x_1)^3}\;.
\end{equation}
When the first steps annihilate we relabel the remaining steps so that
the
positions of the bottom step and anti-step are always $x_1$ and $x_0$.
Our system is symmetric
with respect to the origin, and it is sufficient to consider only the
positive part of the $x$ axis. Symmetry also excludes any flux of
adatoms
between the two bunches, and in the LEM case this implies that the
volumes
of the two bunches are conserved separately.

\subsubsection{Local Mass Exchange Mechanism}

The groove flattening problem is different from the two previous
examples,
where there was a well defined facet with its edge at the position of
the
first step. At the facet edge Eq.\ (\ref{eq:local_diff}) becomes invalid
due
to the vanishing step density. In the groove example, steps annihilate
at
the origin, and there are steps in all regions of space. However, there
could still be points where the density of steps vanishes if the
position of
the second step diverges in the scaling limit. Let us denote the scaled
position of the facet edge by $\xi ^{*}$. Eq.\ (\ref{eq:local_diff}) is
valid only for $\xi >\xi ^{*}$, and the value of $\xi ^{*}$ is unknown a
priori. There are two possible, qualitatively different situations: 1)
$\xi
^{*}>0$, i.e., there is a plateau at the bottom of the groove which grow
as $t^{1/4}$. 2) $\xi ^{*}=0$. There could still be a diverging plateau,
but
it
must grow more slowly than $t^{1/4}$. In what follows we rule out the
first
possibility and show that $\xi ^{*}=0$.

According to Eq.\ (\ref{eq:LEM_nchange}), the number of step
annihilation
events grows with time as $t^{1/4}$. Denoting by $t_n$ the time of the
$n$th
annihilation event, we see that $t_n \sim n^4$ and $t_{n+1}-t_n \sim
n^3$.
We now show that if $\xi^*>0$, the time interval between annihilation
events
is larger than $n^3$. Just before the $n$th annihilation event, the
velocity
of the first step (which must be negative) is given by 
\begin{equation}
\lim_{t \rightarrow
t_n^-}\dot{x}_1=\frac{1}{4}\left(\frac{1}{(x_3-x_2)^3}- 
\frac{2}{x_2^3} \right)\;.  \label{eq:tn_x1dot_lem}
\end{equation}
If $\xi^*>0$, $x_2$ is of order $t_n^{1/4} \sim n$. Eq.\ (\ref
{eq:tn_x1dot_lem}) implies that $x_3-x_2>2^{-1/3}x_2$, and therefore the
distance $x_3-x_2$ is also of order $n$.

After the $n$th annihilation event we relabel the steps. $x_2$ becomes
$x_1$
, $x_3$ becomes $x_2$ and so on. Now the distance between the first two
steps, $x_2-x_1$ is of order $n$. The velocity of the new first step,
which
is maximal (in absolute value) at this time, obeys the following
inequality: 
\begin{equation}
\lim_{t \rightarrow t_n^+}\left|\dot{x}_1\right|<\frac{1}{2(x_2-x_1)^3}
\sim 
{\cal O}\left(n^{-3}\right)\;.
\end{equation}
The step must cross a distance of order $n$ until it annihilates and
therefore $t_{n+1}-t_n$ is at least of order $n^4$, in contradiction
with
the relation $t_{n+1}-t_n \sim n^3$ derived above. Hence $\xi^*=0$.

In order to solve Eq.\ (\ref{eq:local_diff}), we now find two boundary
conditions at $\xi=0$. Consider first the quantity $x_3-x_2$. If $x_2$
diverges
with
time, $x_3-x_2$ must also diverge in order for $\dot{x}_1$ to be
negative
(see Eq.\ (\ref{eq:tn_x1dot_lem})). If $x_2$ does not diverge in the
scaling
limit, $x_3-x_2$ must still diverge in order for the time interval
between
two consecutive annihilation events to diverge in the scaling limit.
Thus, $x_3-x_2$ diverges and the scaling function vanishes at the
origin: 
\begin{equation}
F(0)=0\;.  \label{eq:groove_lem_bc1}
\end{equation}

Another boundary condition is derived from volume conservation in the
$x>0$
half of space. From Eq.\ (\ref{eq:LEM_vchange}) we obtain the condition 
\begin{equation}
\left.
\left[\xi\frac{d}{d\xi}\left(\frac{1}{F}\frac{d^2}{d\xi^2}\frac{3F^2}{2} 
\right)-\frac{1}{F}\frac{d^2}{d\xi^2}\frac{3F^2}{2}\right]
\right|_0=0\;.
\label{eq:groove_lem_bc2}
\end{equation}

\begin{figure}[h]
\epsfxsize=80mm
\centerline{\epsffile{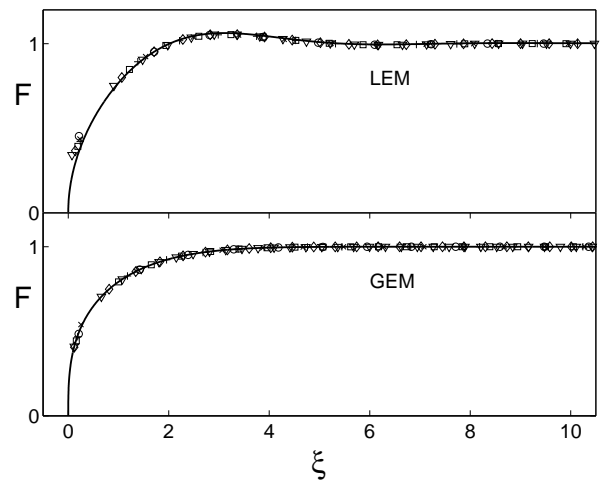}}
\vspace{3mm}
\caption{Solution of the groove flattening scaling function in the LEM 
and GEM cases (solid lines)
compared with scaled density functions from numerical
simulations of the discrete models. Different symbols represent density
functions at
different times.} 
\label{groove_scaling}
\end{figure}

Requiring expansion (\ref{eq:small_F_exp_lem}) to satisfy Eqs.\ (\ref
{eq:local_diff}) and (\ref{eq:groove_lem_bc2}) we are left with two free
expansion coefficients, which we tune in order to satisfy the boundary
condition at infinity. The resulting solution compared to simulation
data is
shown in the upper half of Fig.\ \ref{groove_scaling}.

\subsubsection{Global Mass Exchange Mechanism}

As in the LEM case, we have to find the value of $\xi^*$. From
considerations similar to those used in the LEM case it can be shown
that
steps annihilate fast enough for scaling to occur only if $\xi^*=0$.

In the GEM case the volume is not conserved. However, by summing up the
velocities of all the steps we can calculate the rate of change of the
volume: 
\begin{equation}
\frac{dV}{dt}=-\sum_{n=1}^\infty \dot{x}_n=\frac{1}{2}\lim_{N
\rightarrow
\infty} \frac{1}{(x_N-x_{N-1})^3}\;.  \label{eq:groove_gem_dis_vchange}
\end{equation}
The r.h.s.\ of Eq.\ (\ref{eq:groove_gem_dis_vchange}) is $F_\infty^3/2$,
while the l.h.s.\ can be calculated from Eq.\ (\ref{eq:GEM_vchange})
with $\xi_a=0$, $\xi_b=\infty$ and $\theta^2=t$. This calculation
combined
with
Eq.\ (\ref{eq:groove_gem_dis_vchange}) leads to the boundary condition 
\begin{equation}
\left. \left( F^3-3 \xi F^2F'\right) \right|_0^\infty=F_\infty^3
\;,
\label{eq:groove_gem_bc2}
\end{equation}
which implies 
\begin{equation}
F(0)=0.
\end{equation}

To calculate the scaling function we use the series expansion (\ref
{eq:small_F_exp_gem}) to approximate $F$ near the singular point and
then integrate Eq.\ (\ref{eq:global_diff}) 
numerically from a point where $F$ is
finite. We tune the expansion coefficient $b_1$ to satisfy the boundary
condition at infinity. The resulting solution compared to simulation
data is
shown in the lower half of Fig.\ \ref{groove_scaling}.

\section{Summary and Discussion}

We studied 1D step-flow models for the kinetics of faceting and
relaxation
and showed that the surface profile exhibits scaling behavior. The value
of
the scaling exponents was determined by the mass transport mechanism.
The
scaling functions in all cases considered here were described by the
same
differential equation for the same mass transport mode. These scaling
functions differ from those predicted by Mullins' continuum theory,
which
does not explicitly consider the existence of steps. However, the
scaling
exponents $\gamma $ agree with Mullins' classical theory. This can be
understood as follows. The systems considered here are well below the
roughening temperature of the (low-index) singular surface but large
parts
of the system are rough, with nonzero macroscopic slope and a
differentiable
free energy as a function of orientation. Therefore, if the system shows
scaling behavior at all, the whole system (including the singular
region)
must evolve with the same time dependence as the non-singular part,
which is
accurately described by Mullins' model.

Here we only considered attachment/detachment limited kinetics for local
mass exchange. We have carried out a similar scaling analysis for
diffusion
limited kinetics. Diffusion limited kinetics is also a local mass
exchange
mode and hence the scaling exponents are the same as for
attachment/detachment limited kinetics. However, the
differential equation for 
the diffusion limited scaling function is different from the
attachment/detachment limited case. For diffusion limited
kinetics we found equations, which under the scaling assumption, are
equivalent to the continuum models proposed in Refs.
\onlinecite{OzdemirZangwill,HagerSpohn}.
Our results are valid even for
finite step permeability at sufficiently long times.

Another issue that we would like to investigate further is the condition
for
scaling behavior. When a step profile shows a scaling behavior, the
scaling
exponent should depend on the mass transport mechanism but not the
driving
force of the evolution at the boundary. However, very strong driving
forces
could cause a piling up of steps and destroy the scaling behavior. What
kind
of driving force gives rise to a scaling behavior? For example, for the
flattening of a groove, we used a contact interaction between steps with
opposite signs. What kind of interaction between the steps of opposite
signs
in the middle, in general, results in a profile that obeys scaling? Does
it
depend on the mass transport mechanism? We have examined some artificial
examples (not described here) of interactions between the steps of
opposite
signs that can destroy the scaling behavior but do not know yet the
general
form of the driving forces that admit a scaling ansatz.

One can also consider more general interactions between neighboring
steps in
the non-singular region. Although our scaling analysis is valid for a
general step-step interaction (see Appendix A), we only carried out
simulations
to verify that scaling indeed occurs in the case of simple entropic
and
elastic step-step interactions. What are the interactions that support
scaling? What is the general relationship between the boundary
conditions
and the step interactions that is consistent with a scaling ansatz? If
these
questions can be answered,  we may be able to tell in advance which
surface
systems will show dynamic scaling behavior.

We are grateful to Da-Jiang Liu for helpful discussions. This work was
supported in part by grant No. 95-00268 from the United 
States-Israel Binational Science Foundation (BSF), Jerusalem, Israel, by
the
Korea Research Foundation, Grant D00001, and by the National
Science Foundation (NSF-MRSEC grant \#DMR96-32521).

\appendix
\section{}
In this appendix we give the results of a scaling analysis with a
general
step chemical potential formula. We start by replacing Eq.\ (\ref
{eq:chemical_potential}) with 
\begin{equation}
\mu _{n}^{s}=U\left( x_{n}-x_{n-1}\right) -U\left( x_{n+1}-x_{n}\right)
\;,
\label{eq:general_chemical_potential}
\end{equation}
with $U(x)$ a general analytic function of $x$. Such a formula is
consistent with interactions between nearest neighbor steps.

Next we use this formula together with the step velocities (\ref
{eq:lem_velocity}) and (\ref{eq:gem_velocity}) to rewrite Eq.\ (\ref
{eq:partial2}) in the LEM and GEM cases. Changing the step velocities 
does not alter the expansion (\ref{eq:xisexpansion}) since this
expression is general and depends only on the scaling function $F$. We
can thus use expansion (\ref{eq:xisexpansion}) to expand the new Eq.\
(\ref
{eq:partial2}) in powers of $\theta^{-1}$. 
We find that in the LEM case
\begin{eqnarray}
&\theta&^{\gamma-4}\frac{a^2k}{2}\frac{d^2}{d\xi^2}\left[\frac{1}{F}
\frac{d}{d\xi}\left(\frac{1}{F}\frac{dU\left(\frac{1}{F}\right)}{d\xi}
\right)
\right]\nonumber\\
&-&\frac{\xi}{\gamma}\frac{dF}{d\xi}
+{\cal O}\left(\theta^{\gamma-6}\right)=0\;.  
\label{eq:general_lem_expand_diff}
\end{eqnarray}
This implies that in the LEM case $\gamma=4$ and the differential
equation
for the scaling function is 
\begin{equation}
2a^2k\frac{d^2}{d\xi^2}\left[\frac{1}{F}\frac{d}{d\xi}\left(\frac{1}{F} 
\frac{dU\left(\frac{1}{F}\right)}{d\xi}\right)\right]
-\xi\frac{dF}{d\xi}=0\;.  \label{eq:general_lem_diff}
\end{equation}

In the GEM we find that 
\begin{equation}
\theta^{\gamma-2}a^2k\frac{d^2
U\left(\frac{1}{F}\right)}{d\xi^2}+\frac{\xi}{\gamma}\frac{dF}{d\xi}+
{\cal O}\left(\theta^{\gamma-4}\right)=0\;,
\label{eq:general_gem_expand_diff}
\end{equation}
which implies $\gamma=2$ and 
\begin{equation}
2a^2k\frac{d^2
U\left(\frac{1}{F}\right)}{d\xi^2}+\xi\frac{dF}{d\xi}=0\;.
\label{eq:general_gem_diff}
\end{equation}


\begin{references}
\bibitem{comment_for_the_general_case} In the more general case of a 
finite ratio between the terrace diffusion rate and the step-edge 
attachment rate, $\mu _{n}^{t}$ varies with position and must be 
calculated explicitly by solving the
diffusion equation. This does not change our basic conclusions.
\end{references}
\end{document}